\documentstyle[preprint,pre,aps]{revtex}
\begin{document}
\def\be{\begin{equation}}
\def\ee{\end{equation}}
\def\bc{\begin{center}}
\def\ec{\end{center}}
\def\bea{\begin{eqnarray}}
\def\eea{\end{eqnarray}}

\title{Fluctuation dissipation ratio in the one dimensional
kinetic Ising model}
\author{E. Lippiello and M. Zannetti}
\address { Istituto Nazionale di Fisica della Materia, Unit\`{a}
di Salerno and Dipartimento di Fisica, Universit\`{a} di Salerno,
84081 Baronissi (Salerno), Italy}
\maketitle
\begin{abstract}
The exact relation between the response function $R(t,t^{\prime})$
and the two time correlation function $C(t,t^{\prime})$ is 
derived analytically in the one dimensional kinetic Ising
model subjected to a temperature quench. The fluctuation
dissipation ratio $X(t,t^{\prime})$ is found to depend on time
through $C(t,t^{\prime})$ in the time region where scaling
$C(t,t^{\prime}) = f(t/t^{\prime})$ holds. The crossover from the
nontrivial form $X(C(t,t^{\prime}))$ to $X(t,t^{\prime}) \equiv 1$
takes place as the waiting time $t_w$ is increased from below
to above the equilibration time $t_{eq}$.
\end{abstract}

PACS:  05.70.Ln, 64.75.+g

\noindent e-mail: lippiello@sa.infn.it, zannetti@sa.infn.it

%\begin{multicols}{2}
\section{Introduction}

The time evolution of a system relaxating to equilibrium is
characterized by two distinct time regimes: the off equilibrium
transient for $t< t_{eq}$ and the stationary equilibrium
evolution for $t> t_{eq}$, where $t_{eq}$ is the equilibration time.
Normally, $t_{eq}$ is smaller than experimental times and one can
actually observe equilibration. However, the existence of
situations in which the opposite is true, namely where $t_{eq}$
exceeds by far any practical observation time as in glassy systems
at low temperature, has contributed to arise a lot of interest
in the off equilibrium relaxation regime. It turns out that
quite a useful tool in the investigation of these slowly relaxating 
systems, with and without disorder, is the fluctuation dissipation 
relation.

For definiteness, let us consider a system in equilibrium at some
temperature $T_I$. At the time $t=0$ a quench to a lower temperature
$T_F$ is performed. On top of this primary relaxation process a
second one is activated by switching on an external field at the time
$t_w >0$. Characterizing the response of the system under the
action of the perturbation by the response function $R(t,t^{\prime})$,
the search for a fluctuation dissipation relation aims to connect
$R(t,t^{\prime})$ to the relevant correlation function
$C(t,t^{\prime})$ in the unperturbed relaxation process.
Given $R(t,t^{\prime})$ and $C(t,t^{\prime})$ one can always 
write
\be
R(t,t^{\prime}) = {X(t,t^{\prime}) \over T_F}
{\partial \over \partial t^{\prime}} C(t,t^{\prime})
\label{I1}
\ee
where $t \ge t^{\prime}$. Without any further specification this is
just a definition of the quantity $X(t,t^{\prime})$, which is 
called the fluctuation dissipation ratio (FDR). Eq. (\ref{I1})
acquires predictive power when independent statements are made
about the FDR. Thus, if the shortest time $t^{\prime}$ is
greater than the equilibration time $t_{eq}$, i.e. if one looks
into the time translation invariant equilibrium dynamics, the 
fluctuation dissipation theorem (FDT) requires 
$X(t,t^{\prime}) \equiv 1$. This is no more true when $t^{\prime}
< t_{eq}$. However, if appropriate conditions on 
$t^{\prime}$,$t$,$t_{eq}$ are satisfied it may turn out that
$X(t,t^{\prime})$ depends on the time arguments only through
$C(t,t^{\prime})$. This was first discovered by Cugliandolo
and Kurchan~\cite{Cugliandolo93}
in the context of mean field spin glass models
at low temperature. In that case the system does not equilibrate
$(t_{eq}= \infty)$ and $X(t,t^{\prime}) = X(C(t,t^{\prime}))$
in the asymptotic limit of large times. Subsequently the validity
of this relation was verified for finite dimensional spin glass
models~\cite{Franz95} and also in the coarsening processes of non disordered
systems~\cite{Cugliandolo95,Barrat98}. As a matter of fact
a classification of slowly relaxating systems 
can be made~\cite{Cugliandolo99} on the basis of the behavior of $X(C)$. 

The relation between $X$ and $C$ is important for different reasons. 
From the point of view of analytical calculations it allows to 
close the equations of motion for $R$ and $C$~\cite{Cugliandolo93}. 
From the more
fundamental point of view of the understanding of the off equilibrium
dynamics it can be related to the effective temperature of different
dynamical modes~\cite{Cugliandolo97} 
and under certain hypothesis it provides a connection 
between the relaxation regime and the 
structure of equilibrium states~\cite{Franz98}.

In this paper we analyse the relaxation to equilibrium in the one
dimensional kinetic Ising model quenched from the initial temperature
$T_I$ to the lower final temperature $T_F$. The correlation length
then grows from some initial value $\xi_I$, which we assume
$O(1)$, to the final value
\be
\xi_F = -[\log \tanh (J/T_F)]^{-1}
\label{I2}
\ee
where $J>0$ is the nearest neighbors ferromagnetic interaction.
The equilibration time is defined by
\be
t_{eq}= \xi_F^{2}.
\label{I3}
\ee
By lowering the temperature of the quench $t_{eq}$ can be tuned at will 
with $\lim_{T_F \rightarrow 0} t_{eq} = \infty$ allowing for the
investigation of the slow relaxation coming from the high temperature
side. We compute the response function $R(t,t^{\prime})$ after
switching on a random external field at the time $t_w$ after the
quench. We are then able to analyse in detail the changeover from
the equilibrium to the off equilibrium regime by monitoring the change
in the FDR (or in the integrated response) as $t_w$ is varied from
$t_w > t_{eq}$ to $t_w < t_{eq}$. When $t_w > t_{eq}$ dynamics is
time translation invariant and the usual FDT holds. When the
region $t_w < t_{eq}$ is entered the deviation from FDT occurs. 
However, if the difference between $t_w$ and $t_{eq}$ is sufficiently
large, there is a range of values between $t_w$ and $t_{eq}$ where
$C(t,t^{\prime})$ scales. Whithin this range the FDR depends only
on $C$, while outside there remains an explicit dependence on
$t_w$. In other words with a finite but large $t_{eq}$ the off
equilibrium dynamics follows the pattern of interrupted aging and
we find that the FDR depends only on $C$ as long as aging holds.
The case of the zero temperature quench is the limiting case
$(t_{eq} = \infty)$ where aging occurs for arbitrarily large times
yielding $X(t,t^{\prime})=X(C(t,t^{\prime}))$ at all times.

\section{Unperturbed correlation functions}

In the following we consider a one dimensional ferromagnetic
Ising model with nearest neighbor interaction
\be
{\cal H}[\sigma] = -J \sum_{i=1}^{N} \sigma_i \sigma_{i+1}
\label{II4}
\ee
evolving in time with Glauber single spin flip dynamics
\be
{\partial \over \partial t} P([\sigma],t)  = \sum_i \left\{ w(-\sigma_i)
P([R_{i}\sigma],t) - w(\sigma_i) P([\sigma],t) \right\}
\label{II5}
\ee
where $P([\sigma],t)$ is the probability of realization of the
configuration $[\sigma]$ at the time $t$, $[R_i \sigma]$ is the
configuration with the i-th spin reversed
\be
w(\sigma_i) = {1 \over 2} \left[ 1 - {\gamma \over 2} \sigma_i
(\sigma_{i-1} + \sigma_{i+1}) \right]
\label{II6}
\ee
is the transition rate from $[\sigma]$ to $[R_i \sigma]$
and $\gamma = \tanh ({2J \over T_F})$.

Given an initial probability distribution $P([\sigma],t=0)$,
with the choice (\ref{II6}) for the transition rate the solution 
of (\ref{II5}) for large time reaches the equilibrium Gibbs state
$P_{eq}[\sigma]= {1 \over Z} \exp (-{1 \over T_F} {\cal H}[\sigma])$.
The dynamics in the one dimensional case has 
been solved by Glauber~\cite{Glauber63}.
Let us summarise those properties of the time dependent correlation
functions which will be needed in the following. Assuming that 
the initial state $P([\sigma],t=0)$ is symmetrical 
the probability distribution $P([\sigma],t)$
remains symmetrical throughout yielding $<\sigma_{i}(t)> \equiv 0$
for all time. The equal time and the two time correlation 
functions then are defined by
\be
D_{ij}(t) = \sum_{[\sigma]} \sigma_i \sigma_j P([\sigma],t)
\label{II7}
\ee
\be
C_{ij}(t,t^{\prime}) = \sum_{[\sigma][\sigma^{\prime}]} 
\sigma_i \sigma_j^{\prime} P([\sigma^{\prime}],t^{\prime})
P([\sigma^{\prime}],t^{\prime} \mid [\sigma],t)
\label{II8}
\ee
with $t \ge t^{\prime}$ and $C_{ij}(t,t)=D_{ij}(t)$.
$P([\sigma^{\prime}],t^{\prime} \mid [\sigma],t)$ is the
conditional probability to find the system in the configuration
$[\sigma]$ at the time $t$, given that it was in the configuration
$[\sigma^{\prime}]$ at the earlier time $t^{\prime}$. We assume
that space translation invariance holds at all times.

From (\ref{II5}) and (\ref{II7}) one can show that the
equal time correlation function satisfies the equation of motion
\be
{d \over dt} D_{ij}(t)  = -2 D_{ij}(t) + {\gamma \over 2}
\left[ D_{i,j-1}(t) + D_{i,j+1}(t) + D_{i-1,j}(t) + D_{i+1,j}(t) \right]
\label{II9}
\ee
for $i \neq j$ and ${d \over dt} D_{ii}(t) =0$, since 
$D_{ii}(t)=1$.
Similarly, from (\ref{II8}) taking into account that also the
conditional probability satisfies the master equation (\ref{II5})
one finds
\be
{\partial \over \partial t} C_{ij}(t)  = 
- C_{ij}(t,t^{\prime}) + {\gamma \over 2}
\left[ C_{i-1,j}(t,t^{\prime}) + C_{i+1,j}(t,t^{\prime}) \right].
\label{II10}
\ee
The single spin conditional expectation can be computed 
explicitely obtaining~\cite{Glauber63}
\be
\sum_{[\sigma]} P([\sigma^{\prime}],t^{\prime} \mid [\sigma],t)
\sigma_i = \sum_l \sigma^{\prime}_l F_{i-l}(t-t^{\prime})
\label{II11}
\ee
where $F_{i-m}(t-t^{\prime})= e^{-(t-t^{\prime})}
I_{i-m}(\gamma (t-t^{\prime}))$ and $I_{n}(x)$ are the Bessel
functions of imaginary argument. Then, using the definitions (\ref{II7})
and (\ref{II8}), the two times and the equal times correlation
functions are related by
\be
C_{ij}(t,t^{\prime}) = \sum_l D_{jl}(t^{\prime}) F_{i-l}(t-t^{\prime})
\label{II12}
\ee
or in Fourier space
one finds
\be
C_{k}(t,t^{\prime}) = D_{k}(t^{\prime}) e^{-\gamma_{k}(t-t^{\prime})}
\label{2.10}
\ee
with $\gamma_k =1- \gamma \cos k$.
Given the initial condition $D_{i,j}(0)$ Eq.(\ref{II9}) can be 
solved exactly~\cite{Glauber63}. After some microscopic time
$t_0$, which we assume much smaller than $t_{eq}$, memory of
the initial condition is lost and for $k \ll 1$, $\xi_F \gg 1$
one has~\cite{Bray89}
\be
D_{k}(t^{\prime})= 2({t^{\prime} \over \pi})^{1/2}
{1 \over k^2 + \xi_{F}^{-2}} \int_{0}^{1} dy y^{-1/2} 
[\xi_{F}^{-2} e^{-\xi_{F}^{-2} t^{\prime} y} +
k^{2} e^{(-[k^2 + \xi_{F}^{-2}] t^{\prime} + k^{2} t^{\prime} y)}]
\label{2.13}
\ee
where we have expanded $\gamma_{k}$ to lowest order in $k$ and
$\xi_F^{-1}$ using $\gamma = 1/\cosh(\xi_{F}^{-1})$. The form
(\ref{2.13}) for the equal time structure factor obeys the scaling 
relation $D_{k}(t^{\prime}) =(t^{\prime})^{1/2} g(k^{2}t^{\prime},
t^{\prime}/t_{eq})$ with the limits
\be
D_k(t^{\prime}) \sim \left\{ \begin{array}{ll}
(t_{eq})^{1/2} g_{eq}(k^2 t_{eq}) & \mbox{ for $t^{\prime}/t_{eq} \gg 1$} \\
(t^{\prime})^{1/2}g_{sc}(k^2 t^{\prime}) & \mbox{ for $t^{\prime}/t_{eq} 
\ll 1$}.
\end{array}
\right.
\label{202}
\ee
Inserting (\ref{2.13}) into (\ref{2.10}) and inverting the Fourier transform, 
the corresponding scaling form for the two time correlation
function is obtained
$C_{i,j}(t,t^{\prime}) = 
f \left( {\mid i-j \mid \over (t^{\prime})^{1/2}},
{t \over t^{\prime}},{t^{\prime} \over t_{eq}} \right)$ with
the limiting behaviors
\be
C_{i,j}(t,t^{\prime}) \sim \left\{ \begin{array}{ll}
f_{eq} \left( {\mid i-j \mid \over (t_{eq})^{1/2}},
{t-t^{\prime} \over t_{eq}} \right) & \mbox{for $t^{\prime}/t_{eq} 
\gg 1$} \\
f_{sc} \left( {\mid i-j \mid \over (t^{\prime})^{1/2}},
{t \over t^{\prime}} \right) & \mbox{for  $t^{\prime}/t_{eq} \ll 1$}.
\end{array}
\right.
\label{204}
\ee
The use of the small $k$ approximation (\ref{2.13}) is justified
since for small enough $T_F$ the structure factor builds up a
large and narrow peak about $k=0$ which gives the main contribution
to the integral over $k$. In the following it will be sufficient to consider
the autocorrelation function $(i=j)$. In the case of the zero temperature
quench the scaling behavior at the bottom of (\ref{204}) is obeyed
for all times since $t_{eq}=\infty$ and the explicit form of the
scaling function is given by~\cite{Bray89,Prados97}
\be
C_{i,i}(t,t^{\prime}) = f_{sc}(t/t^{\prime}) =  {2 \over \pi} \arcsin 
\sqrt{{2  \over 1+ {t \over t^{\prime}}}}.
\label{2.17}
\ee

With $T_F>0$ and $t_{eq} < \infty$, $C_{i,i}(t,t^{\prime})$ can
be computed by a combination of analytical and numerical
integration (Appendix). In Fig.1 we have plotted 
$\log C_{i,i}(t,t^{\prime})$ against $x=({t \over t^{\prime}} -1)$
for different values of $\tau = t^{\prime}/t_{eq}$ illustrating
the crossover from the scaling form (\ref{2.17}) to the exponential 
decay corrasponding to  the top of (\ref{204}) as $\tau$ grows 
from small to large values.

\section{Response function}

As stated in the Introduction, let us now assume that after the 
quench to $T_F$, at some time $t_w >0$, a site and time dependent 
external field $h_{i}(t)$ is switched on. We are interested in the 
response in the magnetization to the action of this field.
More specifically, we wish to investigate the relation
between the magnetization response and the correlation function
in absence of the field. 

For sufficiently small external field, the response in the 
magnetization at site $i$ and $t>t_w$ is given by linear response
theory
\be
\Delta <\sigma_{i}(t)> = <\sigma_{i}(t)>_{h} - <\sigma_{i}(t)>_{h=0}
= \sum_{j} \int_{t_{w}}^{t} dt^{\prime} R_{i,j}(t,t^{\prime})
h_{j}(t^{\prime})
\label{3.1}
\ee
where
\be
R_{i,j}(t,t^{\prime}) = \left. {\delta <\sigma_{i}(t)>_{h} \over
\delta h_{j}(t^{\prime})} \right)_{h=0}
\label{3.2}
\ee
is the causal response function. 
The difference
between the two expectation values in (\ref{3.1}) is given by
$\Delta <\sigma_{i}(t)> = \sum_{[\sigma]} \sigma_i \Delta 
P([\sigma],t)$ where $\Delta P([\sigma],t) = P_{h}([\sigma],t) - 
P([\sigma],t)$ is the difference between the probabilities with
and without the field. The time evolution of $P_{h}([\sigma],t)$
is given by the master equaltion (\ref{II5}) with the transition
rate (\ref{II6}) replaced by
$w_{h}(\sigma_i) = w(\sigma_i) + \Delta w(\sigma_i)$ where
$\Delta w(\sigma_i) = - \tanh ({h_i \over T_F}) \sigma_i
w(\sigma_i)$. Taking $h_i / T_F$ sufficiently small
$\tanh ({h_i \over T_F}) \simeq {h_i \over T_F}$ and following
~\cite{Glauber63} up to first order we have
\begin{eqnarray}
\Delta P([\sigma],t) & = &  {1 \over T_F} \sum_{ [\sigma^{\prime}] }
\sum_i \sigma^{\prime}_i  \int_{t_w}^{t} dt^{\prime} h_{i}(t^{\prime})
\left[ w(\sigma^{\prime}_i) P([\sigma^{\prime}],t^{\prime}) + \right. 
\nonumber \\
&  &  \left. w(-\sigma^{\prime}_i) P([R_{i}\sigma^{\prime}],t^{\prime}) 
\right] P([\sigma^{\prime}],t^{\prime} \mid [\sigma],t)
\label{3.3}
\end{eqnarray}
yielding
\begin{eqnarray}
R_{i,j}(t,t^{\prime}) & = &  {1 \over T_F} 
\sum_{[\sigma] [\sigma^{\prime}]} \sigma^{\prime}_j
\left[ w(\sigma^{\prime}_j) P([\sigma^{\prime}],t^{\prime}) +
w(-\sigma^{\prime}_j) P([R_{j} \sigma^{\prime}],t^{\prime}) \right] 
\nonumber \\
&  & P([\sigma^{\prime}],t^{\prime} \mid [\sigma],t) \sigma_i .
\label{3.4}
\end{eqnarray}
Performing the sum over $[\sigma]$ first and using (\ref{II11}) we find
\begin{eqnarray}
R_{i,j}(t,t^{\prime}) & = &  {1 \over T_F} 
\sum_{[\sigma^{\prime}]} \sum_{l} \sigma^{\prime}_j \sigma^{\prime}_l
\left[ w(\sigma^{\prime}_j) P([\sigma^{\prime}],t^{\prime}) + \right. 
\nonumber \\
&  & \left. w(-\sigma^{\prime}_j) P([R_{j} \sigma^{\prime}],t^{\prime}) 
\right] F_{i-l} (t-t^{\prime})
\label{3.5}
\end{eqnarray}
and since only the term with $l=j$ survives in the summation we have
\be
R_{i,j}(t,t^{\prime}) = {1 \over T_F} \left[ 1 - {\gamma \over 2}
(D_{j,j-1}(t^{\prime}) + D_{j,j+1}(t^{\prime})) \right] F_{i-j}(t-t^{\prime}).
\label{3.6}
\ee

In order to recast this result in terms of 
$C_{i,j}(t,t^{\prime})$ let us differentiate (\ref{II12}) with
respect to the time arguments
\begin{eqnarray}
{\partial \over \partial t^{\prime}}  C_{i,j}(t,t^{\prime})   & = &
\sum_l {d D_{j,l}(t^{\prime}) \over dt^{\prime}} F_{i-l}(t-t^{\prime}) +
\nonumber \\
&  & \sum_l D_{j,l}(t^{\prime}) {d \over dt^{\prime}} F_{i-l}(t-t^{\prime})
\label{3.7}
\end{eqnarray}
\be
{\partial \over \partial t} C_{i,j}(t,t^{\prime})  =
-\sum_l D_{j,l}(t^{\prime}) {d \over dt^{\prime}} F_{i-l}(t-t^{\prime})
\label{3.8}
\ee
Adding (\ref{3.7}) and (\ref{3.8}) the summation in (\ref{3.8})
cancels the second one in (\ref{3.7}) yielding
\be
{\partial \over \partial t^{\prime}} C_{i,j}(t,t^{\prime})  +
{\partial \over \partial t} C_{i,j}(t,t^{\prime})  =
\sum_l {dD_{j,l}(t^{\prime}) \over dt^{\prime}}  F_{i-l}(t-t^{\prime}).
\label{3.9}
\ee
Next, using (\ref{II9}) for the time derivative when $l \neq j$,
adding and subtracting a similar contribution with $l = j$ and
using translational invariance we can rewrite
\begin{eqnarray}
{\partial \over \partial t^{\prime}} C_{i,j}(t,t^{\prime})  +
{\partial \over \partial t} C_{i,j}(t,t^{\prime})  & = &
-2 \sum_l \left\{ D_{j,l}(t^{\prime}) -{\gamma \over 2}
[D_{j,l+1}(t^{\prime}) + D_{j,l-1}(t^{\prime})] \right\} F_{i-l}(t- t^{\prime})
+ \nonumber \\
&  &  2 \left\{ D_{j,j}(t^{\prime}) -{\gamma \over 2}
[D_{j,j+1}(t^{\prime}) + D_{j,j-1}(t^{\prime})] \right\} F_{i-j}(t- t^{\prime}).
\label{3.10}
\end{eqnarray}
Using (\ref{II12}) and (\ref{II10}) the first sum in the right hand
side is given by $2 {\partial \over \partial t} C_{i,j}(t,t^{\prime})$,
while the second term coincides, up to a constant factor, with the
right hand side of (\ref{3.6}), since $D_{j,j}(t^{\prime})=1$.
Therefore, we finally get the following expression for the
response function
\be
R_{i,j}(t,t^{\prime})= {1 \over 2T_F}
\left[ {\partial \over \partial t^{\prime}} C_{i,j}(t,t^{\prime})  -
{\partial \over \partial t} C_{i,j}(t,t^{\prime}) \right] 
\label{3.11}
\ee
which can be rewritten in the form (\ref{I1})
\be
R_{i,j}(t,t^{\prime})= {X_{i,j}(t,t^{\prime}) \over T_F} 
{\partial \over \partial t^{\prime}} C_{i,j}(t,t^{\prime})  
\label{3.12}
\ee
with
\be
X_{i,j}(t,t^{\prime})= {1 \over 2}
\left[ 1 - {{\partial \over \partial t} C_{i,j}(t,t^{\prime}) 
\over {\partial \over \partial t^{\prime}} C_{i,j}(t,t^{\prime})}
\right]. 
\label{3.13}
\ee
Alternatively, we can expose 
the deviation from FDT through an additive term
\be
R_{i,j}(t,t^{\prime})= {1 \over T_F}
{\partial \over \partial t^{\prime}} C_{i,j}(t,t^{\prime})  -
{1 \over 2T_F} B_{i,j}(t,t^{\prime})
\label{3.14}
\ee
with 
\be
B_{i,j}(t,t^{\prime}) =
{\partial \over \partial t^{\prime}} C_{i,j}(t,t^{\prime})  +
{\partial \over \partial t} C_{i,j}(t,t^{\prime}) .
\label{3.14}
\ee
When time translation invariance holds we have either $X_{i,j}
(t,t^{\prime}) = 1$ or $B_{i,j}(t,t^{\prime}) = 0$ and the
usual FDT is recovered.

\section{Random external field}

In the simulations~\cite{Barrat98} the external field is taken
random with site independent bimodal distribution 
\be
P[h] = \Pi_i \left[ {1 \over 2} \delta (h_i - h) +
{1 \over 2} \delta (h_i + h) \right].
\label{4.1}
\ee
The reason for this choice is not to bias the evolution toward
the formation of predominantly positive or negative domains
through the introduction of the external perturbation. The
quantity of interest then is the staggered magnetization
\be
M(t,t_w) = {1 \over N} \sum_i \overline{ \Delta <\sigma_{i}
(t)> {h_i \over h}}
\label{4.2}
\ee
where the bar represents the average over the field configurations.
From (\ref{3.1}) and (\ref{4.1}) the integrated response is given by
\be
\chi_{ii}(t,t_w) = {1 \over h} M(t,t_w) = \int_{t_w}^t dt^{\prime}
R_{ii}(t,t^{\prime}).
\label{4.3}
\ee

Dropping the double index and inserting the form (\ref{3.12}) of
the response function, if the FDR depends on time only through
$C(t,t^{\prime})$ we have
\be
T_F \chi(C(t,t_w))= \int_{C(t,t_w)}^{1} dC X(C)
\label{101}
\ee
namely also the integrated response depends on time only through
the autocorrelation function. This occurs when FDT holds with 
$X(C)=1$ yielding
\be
T_F \chi(C(t,t_w)) = \left[ 1 - C(t,t_w) \right] 
\label{102}
\ee
and when the scaling form (\ref{2.17}) holds.
In that case from (\ref{3.13}) follows
\be
X(t,t^{\prime}) = {1 \over 2} \left[ 1+ {t^{\prime} \over t}
\right]
\label{4.6}
\ee
and inverting (\ref{2.17})
\be
{t^{\prime} \over t} = { \sin^{2} \left({\pi \over 2}
C(t,t^{\prime})\right) \over 2 - \sin^{2} \left({\pi \over 2}
C(t,t^{\prime})\right) }
\label{4.7}
\ee
we find
\be
X(C) = {1 \over 2 - \sin^{2} \left({\pi \over 2}
C \right)}.
\label{4.8}
\ee
Inserting this into (\ref{101}) we obtain
\be
T_F \chi(C(t,t_w))  = 
{\sqrt{2} \over \pi} \arctan \left[ \sqrt{2}
\cot ({\pi \over 2} C(t,t_w)) \right] .
\label{4.9}
\ee

We have then proceeded to compute (Appendix) $T_F \chi(t,t_w)$
with the values of the parameters $(t_{eq}=10^3)$
corresponding to the behavior of the autocorrelation function
displayed in Fig.1 and we have plotted $T_F \chi(t,t_w)$
against $C(t,t_w)$ in Fig.2 for different values of $t_w/t_{eq}$.
In order to understand the plot notice that if $t_{eq}$ is finite,
from (\ref{4.3}) follows $\lim_{t \rightarrow \infty} T_F 
\chi(t,t_w) = {T_F \over h}M_{eq} =1$ where $M_{eq}$ is the 
equilibrium value of the magnetization. On the other hand,
with a finite $t_{eq}$ one has also $\lim_{t \rightarrow \infty}  
C(t,t_w) = 0$. Therefore, when plotting $T_F\chi$ vs. $C$
all the curves starting out at $(C=1,T_F\chi=0)$ must end up
in the same point $(C=0,T_F\chi=1)$. The dependence on
$t_w/t_{eq}$ enters on how the initial and the final point
are joined. Thus, if $t_w/t_{eq}>1$, FDT holds over the entire
time interval $(t_w,t)$ and the plot is linear according to
(\ref{102}). However, if $t_w/t_{eq}<1$ then it is possible to
have also $t/t_{eq}<1$. In that case $C(t,t^{\prime})$ obeys 
the scaling form (\ref{2.17}) and in the range of values
of $C$ where this holds, $T_F\chi$ follows the shape
(\ref{4.9}). This forces the plot to fall below the straight 
line of the FDT, but eventually as $C$ decreases the plot must 
raise again in order to reach the value $T_F\chi=1$ at $C=0$.
Therefore, the peculiar shape of the curves displaying a 
change in concavity is a consequence of a finite 
equilibration time. The final upword bending of the curves
corresponds to interrupted aging and that is where the
curves do depend on $t_w$. Furthermore, the range of values
where the plot follows the shape (\ref{4.9}) is larger the smaller
is the value of $t_w/t_{eq}$. In the limiting case
$t_{eq}=\infty$ aging holds for all time and the plot obeys
(\ref{4.9}) over the entire range of $C$ values.

\section{Conclusions}

The relaxation dynamics of the one dimensional Ising model
allows to analyse in detail the transition from the off
equilibrium to the equilibrium regime. In particular,
we have obtained the crossover in the FDR from the nontrivial
form $X(C)$ given by (\ref{4.8}) to $X(t,t^{\prime}) \equiv 1$
as a manifestation at the level of the response function
of the crossover in the underlying correlation function
from aging to time translation invariance.

A comment should be made about the shape of $X(C)$. In the case
of the zero temperature quench (\ref{4.8}) holds for all time.
On the other hand, the zero temperature quench is a phase
ordering process eventually leading to the coexistence of
ordered phases as in the quench below the critical point
of a system with a finite critical temperature. In the latter case
$X(C)$ displays~\cite{Barrat98,Cugliandolo99} a qualitatively
different behavior decreasing from $1$ and flattening to zero, 
while in our case $X(C)$ decreases from $1$ toward $1/2$ as $C$ 
goes to zero. Although we do not have a complete understanding
of the origin of this discrepancy, we believe this to be related
to the absence in the one dimensional case of the asymmetry term
in the relation between $R(t,t^{\prime})$ and $C(t,t^{\prime})$.
In the context of Langevin dynamics one can derive~\cite{Cugliandolo94}
in full generality 
\be
R(t,t^{\prime}) = {1 \over 2T_F} \left( {\partial \over \partial t^{\prime}}
- {\partial \over \partial t} \right) C(t,t^{\prime})
-{1 \over 2T_F} A(t,t^{\prime})
\label{301}
\ee
where $A(t,t^{\prime})$ is the asymmetry term which vanishes for linear
dynamics. From (\ref{301}) the FDR takes the following general
form
\be
X(t,t^{\prime}) = {1 \over 2} \left[ 1 -
{{\partial \over \partial t}C(t,t^{\prime}) \over 
{\partial \over \partial t^{\prime}}C(t,t^{\prime})} \right]
- {1 \over 2} {A(t,t^{\prime}) \over
{\partial \over \partial t^{\prime}}C(t,t^{\prime})}
\label{302}
\ee
and if we assume scaling $C(t,t^{\prime}) = f(t/t^{\prime})$
the square brackets contribution is given by (\ref{4.6})
independently from the form of $f(x)$.
Therefore, if we accept that $X$ is a function of $C$ when
scaling holds, in order to have $\lim_{C \rightarrow 0}
X(C) < 1/2$ the asymmetry term must necessarily be nonzero.
Now, from Eq.(\ref{3.11}) follows that in the one dimensional
Ising model with Glauber dynamics the asymmetry is absent.
Indeed, in this case as Eq.s (\ref{II9}) and (\ref{II10}) show, dynamics
is linear. Another example of linear dynamics leading to
$\lim_{C \rightarrow 0} X(C) = 1/2$ is the massless gaussian
model~\cite{Cugliandolo94}. Conversely, if one considers
the Ising model with higher dimensionality and a finite critical
temperature, the equations of motion for the two point
correlation functions are coupled to higher correlation
functions producing a nonlinearity which in turn is expected
to produce a nonvanishing asymmetry in the off equilibrium
regime.

\section{Appendix}

In order to carry out the computation of $T_F\chi(t,t_w)$ we start from
the sum of the time derivatives of $C_k(t,t^{\prime})$
obtained from (\ref{2.10})
\be
\left({\partial \over \partial t^{\prime}} +
{\partial \over \partial t} \right) C_k(t,t^{\prime}) =
{d D_k(t^{\prime}) \over dt^{\prime}} e^{-\gamma_k (t-t^{\prime})}.
\label{A1}
\ee
Using (\ref{2.13}) for $D_k(t^{\prime})$
and carrying out integrations by parts we find
\be
{d D_k(t^{\prime}) \over dt^{\prime}} = {2 \over \sqrt{\pi t^{\prime}}}
e^{-\xi_F^{-2}t^{\prime}} 
-2\sqrt{{t^{\prime} \over \pi}} e^{-(k^2+\xi_F^{-2})t^{\prime}}
k^2 \int_0^1 {dy \over \sqrt{y}} e^{k^2 t^{\prime}y}.
\label{B1}
\ee
Inserting this result into (\ref{A1}) and integrating over $k$
we have
\begin{eqnarray}
B(t,t^{\prime}) & = &
\left({\partial \over \partial t^{\prime}} +
{\partial \over \partial t} \right) C(t,t^{\prime}) =
\int_{-\infty}^{\infty} {dk \over 2\pi} {d D_k(t^{\prime}) \over dt^{\prime}}
e^{-{1 \over 2}(k^2+\xi_F^{-2})(t-t^{\prime})} = \nonumber \\
& & {1 \over \pi} \sqrt{{2 \over t^{\prime}(t - t^{\prime})}}
e^{-{1 \over 2}\xi_F^{-2}(t+t^{\prime})} -
{1 \over \pi} e^{-{1 \over 2}\xi_F^{-2}(t+t^{\prime})}
\int_0^1 dy \sqrt{{2 t^{\prime} \over y(t+t^{\prime} -2yt^{\prime})^3}}
= \nonumber \\ 
& & {1 \over \pi (t+t^{\prime})} \sqrt{{2(t-t^{\prime}) \over t^{\prime}}}
e^{-{1 \over 2}\xi_F^{-2}(t+t^{\prime})}.
\label{A2}
\end{eqnarray}

Next, Fourier transforming the equation of motion (\ref{II9}) 
one obtains 
\be
{d \over dt}D_k(t) = -2 \gamma_k D_k(t) +r(t)
\label{A3}
\ee
with
\be
r(t) =  {e^{-{1 \over 2} \xi_F^{-2}t} \over \sqrt{\pi t}}
+  \xi_F^{-1} Erf(\sqrt{\xi_F^{-2}t}) 
\label{A4}
\ee
where $Erf$ is the error function. Inserting (\ref{A3}) in
the right hand side of (\ref{A1}) and using ${\partial \over
\partial t} C_k(t,t^{\prime}) = -\gamma_k C_k(t,t^{\prime})$ 
one finds
\be
\left({\partial \over \partial t^{\prime}} -
{\partial \over \partial t} \right) C_k(t,t^{\prime}) =
r(t^{\prime}) e^{-{1 \over 2}(k^2+\xi_F^{-2}) (t-t^{\prime})}
\label{A5}
\ee
which after integration over $k$ gives
\be
\left({\partial \over \partial t^{\prime}} -
{\partial \over \partial t} \right) C(t,t^{\prime}) =
{2 \over \pi} {e^{-{1 \over 2} \xi_F^{-2}(t+t^{\prime})} \over
\sqrt{2t^{\prime} (t-t^{\prime})}} +
\sqrt{{2 \over \pi}} \xi_F^{-1}
{Erf(\sqrt{\xi_F^{-2} t^{\prime}}) \over
\sqrt{t-t^{\prime}}} e^{-{1 \over 2} \xi_F^{-2}(t-t^{\prime})}.
\label{A6}
\ee
Inserting this result in (\ref{3.11}) with $i=j$ the integrated
response $T_F \chi (t,t_w)$ is obtained carrying out numerically
the integration in (\ref{4.3}). 

Similarly, the autocorrelation function
is obtained by taking the difference of (\ref{A2}) and (\ref{A6})
and carrying out numerically the time integration in
\be
C(t,t_w) = 1 + \int_{t_w}^{t} ds {\partial \over \partial s}
C(s,t_w).
\label{A7}
\ee

Aknowledgements - This work has been partially supported
from the European TMR Network-Fractals c.n. FMRXCT980183 and
from MURST through PRIN-97.

\newpage

Figure Captions

Fig.1 - Plot of the autocorrelation function for different 
values of $\tau = t^{\prime}/t_{eq}$ and $t_{eq}=10^3$. The continuous
line is the plot of the anlytical solution (\ref{2.17}) coresponding to
$T_F=0$.

Fig.2 - Plot of the integrated response for different 
values of $t_w/t_{eq}$ and $t_{eq}=10^3$. The continuous
line is the plot of the anlytical solution (\ref{4.9}) coresponding to
$T_F=0$.

%\end{multicols}
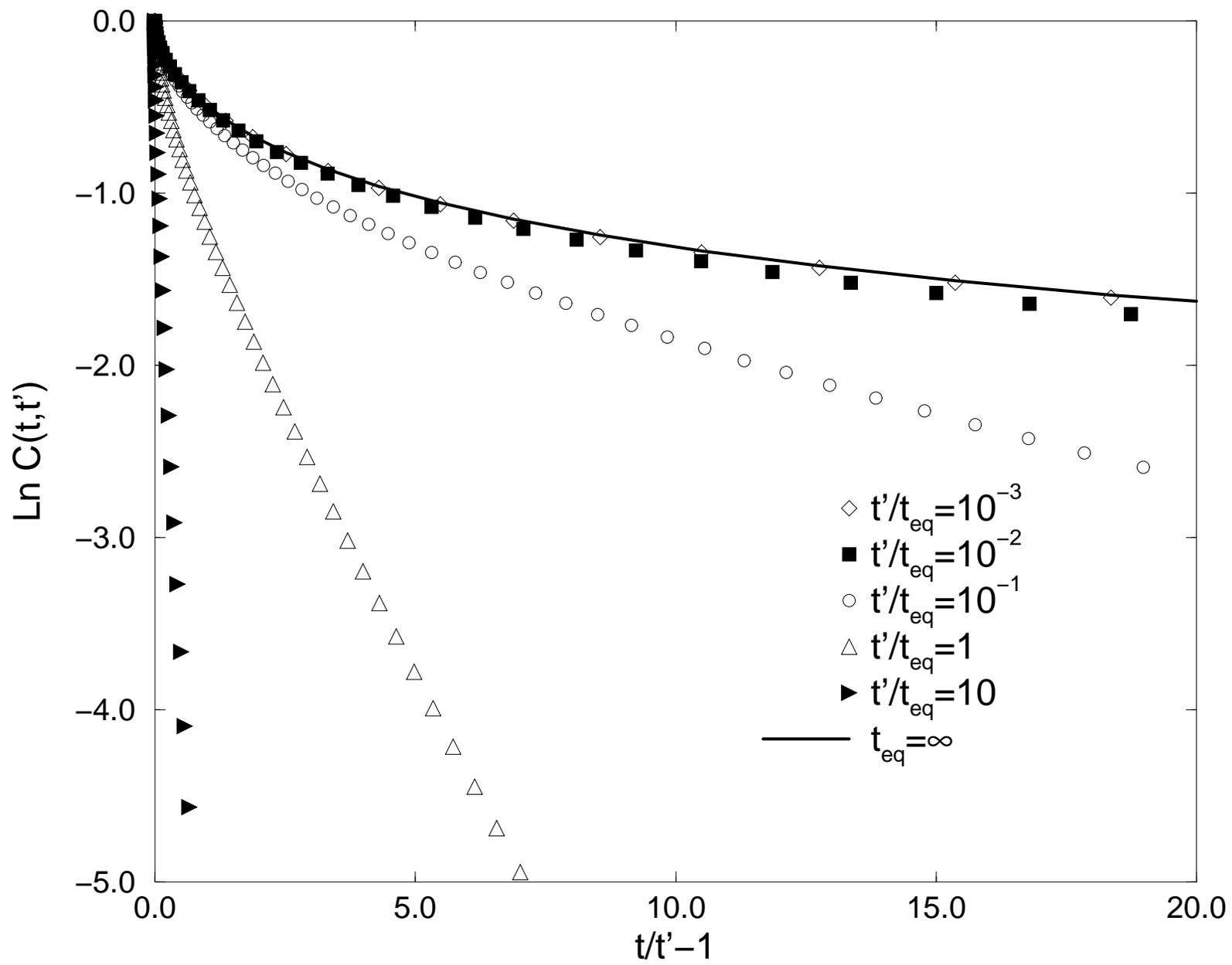
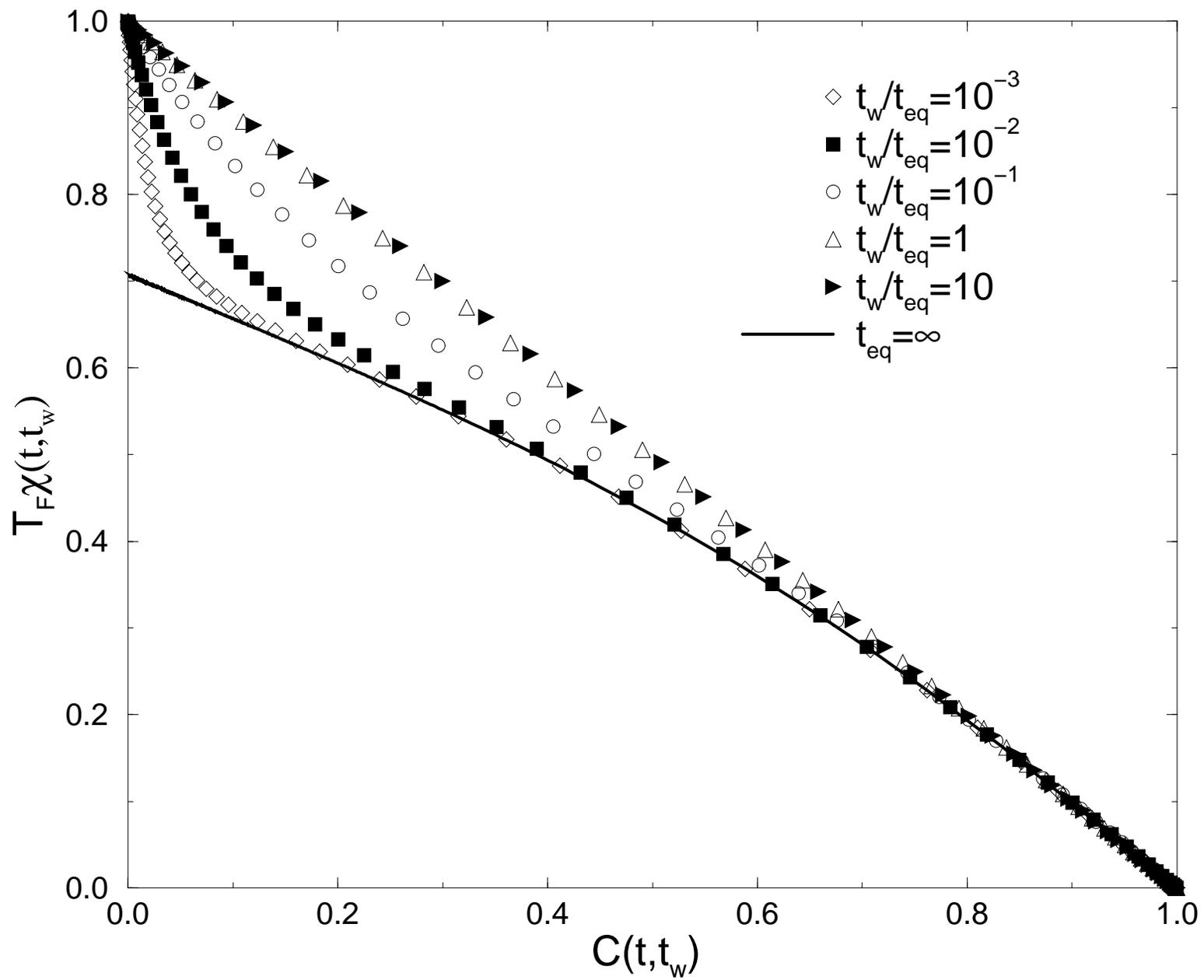
\end{document}